%% file: main.tex
\documentclass[journal,onecolumn]{IEEEtran}

\usepackage{engord}
\usepackage{url,amsmath,amssymb,cite}
\usepackage{subfigure}
\usepackage{color}
\usepackage{amsmath}
\usepackage{amsmath}
\usepackage{graphicx}
\usepackage{graphics}
\usepackage[table]{xcolor} % DVI only
\usepackage{multirow}
\usepackage{rotating}
\usepackage{url}
\usepackage{amssymb}

\definecolor{BgGray}{gray}{0.9}%
\definecolor{BgGray2}{gray}{0.96}%
\definecolor{RowColorOdd}{named}{BgGray2}%
\definecolor{RowColorEven}{named}{white}%
\definecolor{comments}{gray}{.5}

\begin{document}

%\title{IEEE 802.11n kills Opportunistic Routing}
%\title{IEEE 802.11n kills MAC Diversity}
\title{Towards MAC/Anycast Diversity in IEEE 802.11n MIMO Networks}
%{\Large Technical Report}\\
%{\large 02.03.2012}\\
%}

\author{
\IEEEauthorblockN{Anatolij Zubow, Robert Sombrutzki and Markus Scheidgen}\\
\IEEEauthorblockA{Humboldt University Berlin\\
%Rudower Chaussee 25, Berlin, Germany\\
\{zubow,sombrutz,scheidge\}@informatik.hu-berlin.de}\\
}

\maketitle

\begin{abstract}

\emph{Opportunistic Routing} (OR) is a novel routing technique for wireless mesh networks that exploits the broadcast nature of the wireless medium. OR combines frames from multiple receivers and therefore creates a form of \emph{Spatial Diversity}, called \emph{MAC Diversity}~\cite{Miu2005}. The gain from OR is especially high in networks where the majority of links has a high packet loss probability.

The updated IEEE 802.11n standard improves the physical layer with the ability to use multiple transmit and receive antennas, i.e. Multiple-Input and Multiple-Output (MIMO), and therefore already offers spatial diversity on the physical layer, i.e. called \emph{Physical Diversity}, which improves the reliability of a wireless link by reducing its error rate.

In this paper we quantify the gain from MAC diversity as utilized by OR in the presence of PHY diversity as provided by a MIMO system like 802.11n. We experimented with an IEEE 802.11n indoor testbed and analyzed the nature of packet losses. Our experiment results show negligible MAC diversity gains for both interference-prone 2.4\,GHz and interference-free 5\,GHz channels when using 802.11n. This is different to the observations made with single antenna systems based on 802.11b/g~\cite{Miu2005}, as well as in initial studies with 802.11n~\cite{Shrivastava2008}.

\end{abstract}


%\markus{Die implikationen sollten wirklich im nachgang und nicht bereits waehrend der Messung erlaeutert werden. }\tolja{OK;Macht Sinn.}

%% Introduction
\input{introduction}

%% Concrete Research
%\input{motivation}

%% Theory
\input{theory}

%% measurements
\input{evaluation}

%% implications
%\input{discussion}

%% Background and Related Work
\input{related_work}

%% Intro measurements
\input{conclusion}

\bibliographystyle{IEEEtran}
\bibliography{IEEEabrv,mybib}

\end{document}

%% file: introduction.tex
\section{Introduction}\label{sec:introduction}

Modern routing schemes for wireless mesh networks explicitly exploit the broadcast nature of the wireless medium. An unicast packet destined to a specific node is not only received by the intended node, but also by other one-hop neighbors. Traditional routing (also called single-path routing) treats the broadcast nature as a disadvantage, because it induces interference. \emph{Opportunistic Routing} (OR), also called any-path routing, is such a modern broadcast exploiting routing scheme. It dynamically selects from multiple network routes~\cite{Hsu2011} and therefore improves link reliability and overall system throughput. OR creates \emph{Spatial Diversity} (SD) on the MAC layer by combining frames from multiple receivers. This diversity from selecting one out of multiple receivers is also called \emph{MAC Diversity} (MD) or \emph{Anycast Diversity}. Examples of OR protocols are MRD~\cite{Miu2005}, ExOR~\cite{Biswas2005}, McExOR~\cite{Zubow2007} and MORE~\cite{Chachulski2007}. In the past, OR was evaluated in wireless mesh networks with single antenna nodes, i.e. Single-Input Single-Output (SISO), mainly based on the outdated 802.11a/b/g standards~\cite{Biswas2005,Chachulski2007}. By using an OR protocol like MORE the throughput can be doubled compared to state-of-the-art best path routing protocol~\cite{Chachulski2007}.

\begin{figure}[ht]
   \begin{center}
       \includegraphics[width=0.5\linewidth]{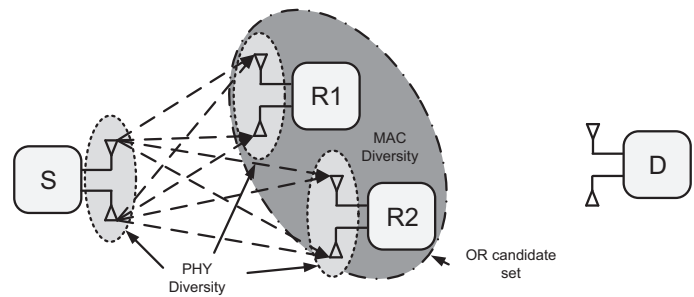}
   \end{center}
   \caption{Combining MAC diversity with PHY diversity.}
   \label{fig:mac_phy_div}
\end{figure}

To benefit from MAC diversity two conditions must be met. First, the majority of operational links must have a high packet loss probability. Second, packet losses among different receivers must be independent or highly uncorrelated.% Finally, make sure that only uncorrelated receivers are grouped together for an OR transmission.

The updated IEEE 802.11n~\cite{IEEE802112009} standard promises faster networks with an increased WiFi coverage. The most important improvement on the PHY layer is the ability to receive and/or transmit simultaneously on multiple antennas (MIMO). The improvements from multiple antennas are two-fold. First, using multiple antennas at the receiver and transmitter side offers a \emph{Spatial Diversity} (SD) gain, also called PHY diversity, which improves the reliability of a wireless link by reducing its error rate. Second, instead of SD MIMO channels can be used to simultaneously transmit multiple data streams through different antennas. This \emph{Spatial Multiplexing} (SM) technique significantly increases the maximum data rate linearly with the number of antennas. % Note, when SM then no SD.
 
%The new updated 802.11n standard improves the physical layer with the ability to use multiple transmit and receive antennas (i.e. MIMO) and therefore already offers spatial diversity on the physical layer, i.e. called \emph{Physical Diversity} (PD). 
An open research question is the combined use of OR and MIMO systems like 802.11n (Fig.~\ref{fig:mac_phy_div}). Spatial diversity obtained at the PHY layer diminishes the adverse effects of signal fading. However, due to the small spacing between antennas, shadowing based channel variations cannot be eliminated. Furthermore, it is hard to combat signal corruption due to interference, e.g. hidden-nodes are common in wireless mesh networks. On the other side, both problems (shadowing and interference) can be eliminated by exploiting macro diversity which is achieved by OR/MD, because the nodes are well spatially separated. Thus current OR research tries to quantify the MAC diversity gain offered by OR in the presence of PHY diversity created by MIMO systems like 802.11n~\cite{Shrivastava2008}. 

In this paper, we measure and analyze packet losses from an 802.11n MIMO-based indoor network and determine possible MAC diversity gain as offered by OR based on the characteristics of the analyzed MIMO links.

The paper is organized as follows. In Section~\ref{sec:understanding_or} we explain how much gain can be expected from OR and what factors have an impact. Next, in Section~\ref{sec:eval} we present experiment results from an 802.11n MIMO based indoor testbed. The results are analyzed and discussed. Thereafter in Section~\ref{sec:relatedwork}, the most important related work is presented and compared with our results. Finally, we conclude our paper and give an outlook.

%% file: theory.tex
\section{Understanding the Gain from OR}\label{sec:understanding_or}

In this section, we explain how much gain can be expected from MAC Diversity (MD) as utilized by OR and what environmental factors have an impact. Note, that the performance gain of OR protocols compared to traditional single-path routing is not exclusively based on MAC diversity and is also related to other aspects. For example, a significant gain of OR protocols comes from Multi-Path Diversity (MPD): the packets of the same flow are routed through multiple paths. This increases the spatial reuse and allows more concurrent transmissions. Furthermore, the use of MPD combined with a MAC protocol like 802.11 results in a medium contention gain. With MPD the probability to access the medium is higher. We focus on the gain from OR achieved through MD only.

\begin{figure*}[t]
  \begin{center}
  \mbox
    {
      {
        \scalebox{1}
        {
            \includegraphics[width=0.35\linewidth]{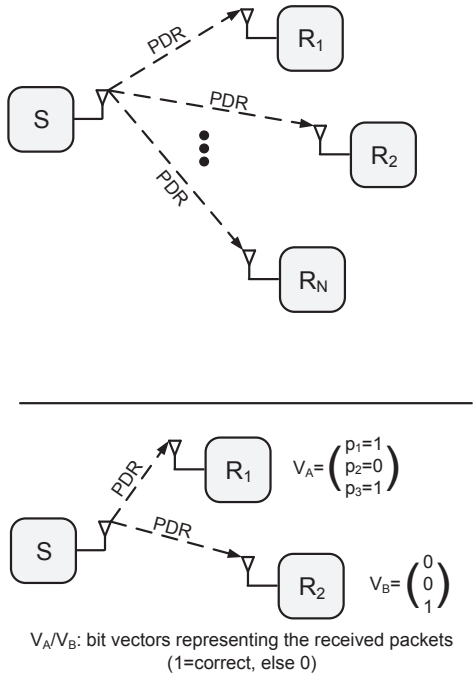}
        }
      }
      {
        \scalebox{1}
        {
            \includegraphics[width=0.50\linewidth]{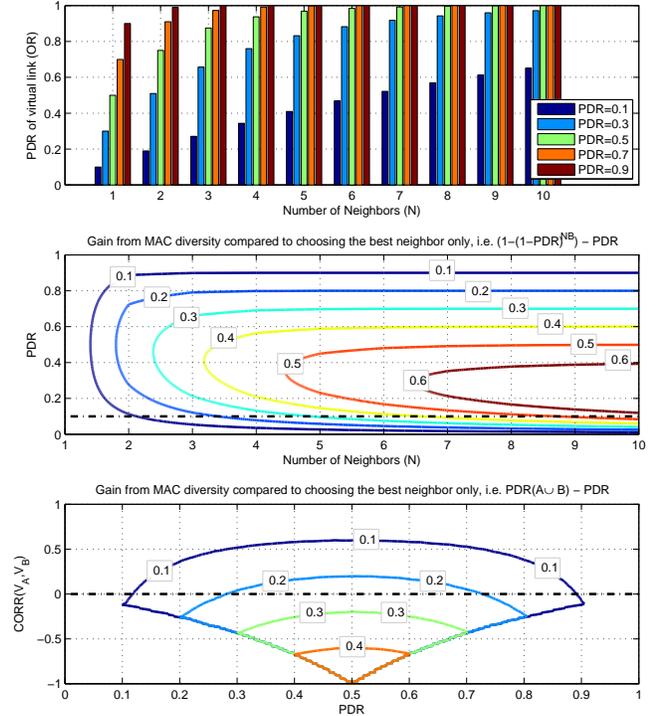}
        }
      }
    }
   \caption{Impact factors on the performance of MAC Diversity: (i) PDR to neighbor nodes, (ii) number of neighbor nodes grouped into candidate set, (iii) correlation of packet losses among different neighbors.}
   \label{fig:theo}
   \end{center}
\end{figure*}

In OR, a single transmitter transmits packets to a candidate relay set. For an OR transmission to be successful it is sufficient that at least one candidate is able to receive the packet (anycast). Therefore, the concept of a virtual link representing the communication link of an OR transmission was introduced. With OR, the packet reception is improved, i.e. the PDR of the virtual link is higher than the PDR of the particular links. Fig.~\ref{fig:theo} (top) shows the impact of the size of candidate relay sets (N) on the PDR of the corresponding virtual link. 

Fig.~\ref{fig:theo} (middle) shows the direct gain from MD as a contour plot. We can observe that the advantage from MD is highest for weak links (low PDR) and for large $N$ (large number of candidates). The gain from MD is low if the PDR of particular links is already high. Thus, in a network with high PDR links, the expected gain from MD is small. Note, that from a practical point of view due to OR coordination overhead the size of the candidate set is mostly restricted (typically 3-5 nodes~\cite{Biswas2005}), and thus very weak links (PDR $\leq 0.1$) cannot be used.

A large number of links with weak PDRs is not a sufficient criterion to benefit from MD. It also depends on independent (or at least highly uncorrelated) packet losses at different receivers.
There is no gain from MD for two receivers with dependent packet losses, i.e. a packet is either received by both receivers or no receiver. So far we have assumed packet loss at different receivers to be independent. Fig.~\ref{fig:theo} (bottom) shows the impact of correlation for two receivers as a contour plot. The correlation, CORR$(V_A,V_B)$, is calculated from bit vectors, where each bit represents whether a packet was received or not. The result indicates that for a fully uncorrelated receiver pair with a PDR of 0.5, the gain from MD is 0.25 (the PDR of the virtual link is 0.75). However, even with a moderate correlation of 0.2 the gain from MD drops below 0.2 (the PDR of the virtual link is 0.7). With a correlation of 0.6 the gain from MD is less than 0.1. Note, that the gain from MD is highest for a negative correlation coefficient, i.e. the probability of packet reception at one receiver is higher when the packet was not successfully received by the other receiver. For two receivers with a PDR of 0.5 each and a correlation coefficient of -1, the PDR of the resulting virtual link is 1.

\begin{figure}[ht]
   \begin{center}
       \includegraphics[width=0.8\linewidth]{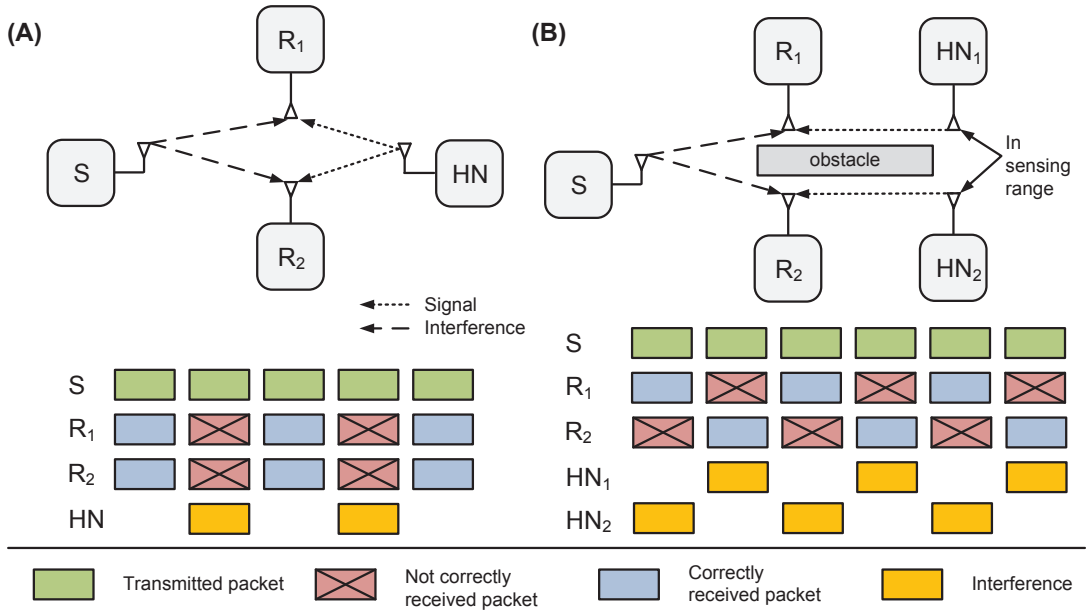}
   \end{center}
   \caption{Illustrative examples for correlated packet losses: (A) $\rho_{R_1,R_2} = 1$ (B) $\rho_{R_1,R_2} = -1$.}
   \label{fig:theo_corr}
\end{figure}

Which environmental factors cause correlated packet losses? Imagine two receivers that are influenced by a single hidden node (Fig.~\ref{fig:theo_corr}, left). Every time when the hidden node transmits, it corrupts the packet reception for both receivers. In this case, the correlation coefficient is 1. Now imagine two receivers that are influenced by two hidden nodes. Further imagine that these hidden nodes sense each other and therefore send alternately. Furthermore, each hidden node only corrupts one receiver. The result is a high correlation with a negative coefficient. In this example, each time one of the receivers correctly receives a packet, the other receiver fails. The correlation coefficient is -1.

From the practical point of view correlation coefficients of $\rho \geq 0$ are more common. To benefit from MD, an OR transmission must have a highly uncorrelated set of relay candidates.

%% file: evaluation.tex
\section{Evaluation}\label{sec:eval}

The goal of this section is to evaluate the nature of packet losses in MIMO 802.11n networks. There are two reasons for packet loss in wireless mesh networks: (i) weak signals and (ii) interference~\cite{Raman2009}. Since we can control interference, i.e. by utilizing an unused channel, will are able to analyze both weak signal and interference based packet losses separately. Moreover, we will study the effect of the used 802.11n MIMO mode; i.e. \emph{Spatial Diversity} (Table~\ref{tab:mcstab}, $\mathrm{MCSIdx} \leq 7$) vs. \emph{Spatial Multiplexing} (Table~\ref{tab:mcstab}, $\mathrm{MCSIdx} \geq 8$).

The rest of this section is structured as follows. At first we present the used experimental methodology like the used 802.11 hardware, the experimental setup and the scenarios to be studied. Thereafter, the experimental results are presented. The implications are discussed in the next section.
%\markus{Diese inhaltsangaben kann man noch ausdehnen: scenarien, analyze linnk laenge, pdr, ...}

%\markus{Ich habe versucht den Sinn neu zu formulieren, bitte pruefen. Eigentlich gehoert das meiner Meinung mit in die Einleitung.}
%MAC layer spatial diversity makes a network less prone to interference based packet losses. The possible spatial diversity gains on MAC layer therefore depend on the existence of lossy links. To analyze the influence of MIMO induced PHY layer spatial diversity on MAC layer diversity gains, we measure MIMO's impact on the statistical dependence of packet losses.

%
% Measurement setup
%
\subsection{Experimental Methodology}
All experiments were conducted in our 802.11n indoor testbed, which is part of the Humboldt Wireless Lab (HWL~\cite{Zubow2011}). The nodes were placed indoors, spanning multiple buildings and floors, as depicted in Fig.~\ref{fig:testbedtopo}. The network has the following characteristics: 20\% of node pairs have a Euclidean distance of less than 10\,m to each other whereas 10\% are separated by more than 45\,m. The median inter-node distance is 22.5\,m.

\vspace{2.5 mm}

\subsubsection*{Wireless Node}
The experiment network consists of 46 Netgear WNDR3700 routers. The WNDR3700 is an off-the-shelf wireless router with an Atheros (AR7161) MIPS CPU, running at 680\,MHz, and 64\,MB of RAM. It has two 802.11n radios; each radio has 4 internal metamaterial antennas from Rayspan\footnote{see http://www.commnexus.org/assets/011/9474.pdf}. The first radio is a dual-band (Atheros AR9220) that can operate in both the 2.4 and the 5\,GHz band, but Atheros restricted the use to the 5\,GHz band. The second radio can only be used in the 2.4\,GHz band (Atheros AR9223). Both radios support 2x2 SM-MIMO channel bonding and can use the short guard OFDM interval (SGI). Both WiFi chips also support STBC to achieve a transmit diversity gain. The optional transmit beamforming is not supported. As driver, we used the open source ath9k developed by the linux-wireless project\cite{linux-wireless}. For more information on our testbed refer to our technical report~\cite{Zubow2011}.

\vspace{2.5 mm}

\subsubsection*{Experiment}
We performed broadcast experiments. Each experiment consists of a sequence of rounds and in each round only one the 46 nodes is transmitting and all others act only as receivers. This ensures that the results are not influenced by internal interference. 
The nodes transmit MAC broadcast packets at a low packet rate to avoid problems like network saturation. The different Modulation and Coding Schemes (MCS) and channel widths (20 and 40\,MHz) were used in a round robin fashion. 

In each round, for each MCS and packet size combination a total of 10,000 packets were transmitted in MAC broadcast mode and the receivers captured the packets using the 802.11 monitor mode. We performed 46 rounds so that each node was able to transmit exclusively. We used the receiver's captured packet traces to analyze the nature of packet losses.\footnote{All experimental results are available as PCAP dump files which can be downloaded from our website: http://hwl.hu-berlin.de/uploads/measurement/or80211n/}.

\vspace{2.5 mm}

\subsubsection*{Scenarios}
We want to understand the nature of packet losses in 802.11n. Especially, we want to determine the environmental factors that influence packet loss. Therefore, we performed three experiments for three different scenarios:
\begin{enumerate}
	\item an occupied (busy) channel from the 2.4\,GHz ISM band
	\item an unused channel from the 5\,GHz ISM band
	\item an unused channel from the 5\,GHz ISM band with artificially induced interference 
\end{enumerate}

\begin{figure}[ht]
   \begin{center}
       \includegraphics[width=0.5\linewidth]{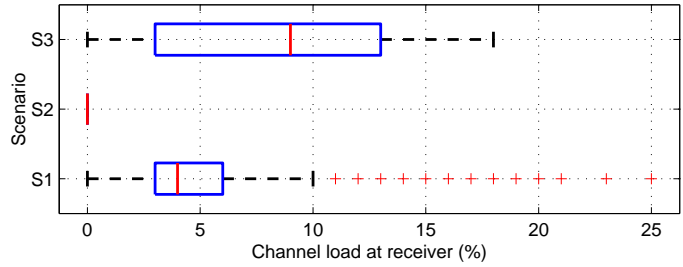}
   \end{center}
   \caption{Channel load as measured at each node before experiment execution.}
   \label{fig:channel_load}
\end{figure}

The focus of the first scenario is to analyze packet losses caused by weak signals or (external) interference. Therefore, we selected channel 6 (2437\,MHz). This channel is used by our campus 802.11 network for serving student's internet traffic. The channel is very busy and even in the night, a significant number of 802.11 beacon frames was observed. Preceding our experiments with that channel, we measured the channel load at each node for 1\,hour.\footnote{The channel load was measured using the hardware registers of the Atheros 802.11n chip. For more information refer to our technical report~\cite{Zubow2011}.} The results of this measurement are shown in Fig.~\ref{fig:channel_load}. We can see that the channel load depends heavily on the spatial location of the node, i.e. it can range from as low as 0 to as high as 31\%, with a median of 4\%.

For the second scenario we aimed to analyze packet losses caused by weak signals only. We selected an unused channel, i.e. channel 161 (5805\,MHz). The public use of this channel is prohibited by German regulations. Preceding measurements of channel load showed zero load at all times. We therefore assume no impact from external interference on this channel (Fig.~\ref{fig:channel_load}).

\begin{figure}[ht]
   \begin{center}
       \includegraphics[width=0.7\linewidth]{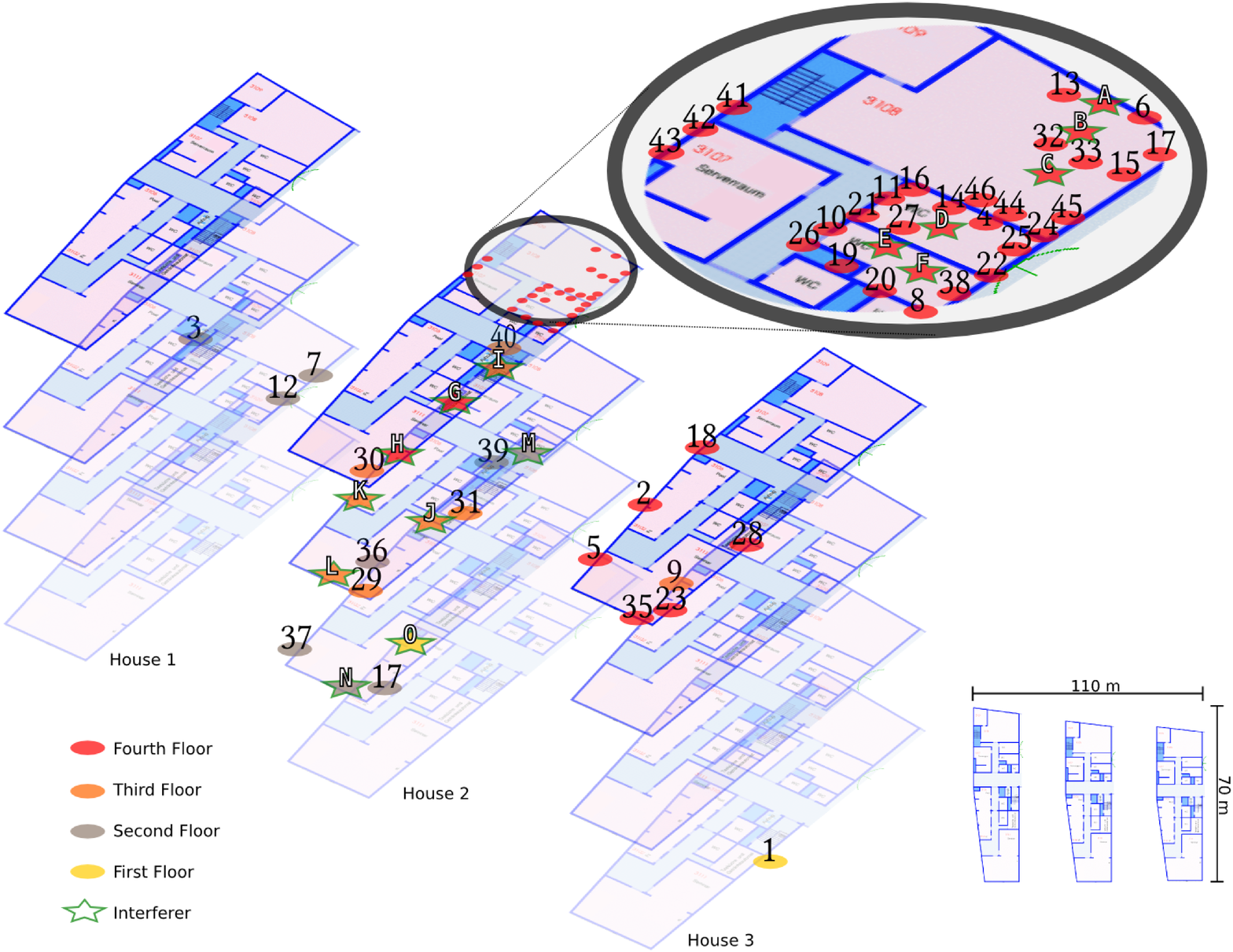}
   \end{center}
   \caption{Testbed topology - 46 nodes (1-46) were distributed indoors among 3 buildings on 4 floors. Furthermore, the location of the 15 interferer nodes (A-O) used in scenario 3 is shown.}
   \label{fig:testbedtopo}
\end{figure}

Finally, the third scenario is different from the first scenario: we induced interference ourselves to control the amount of interference. We selected the empty channel 161 again. This time, 15 additional interferer nodes (A-O) where placed as illustrated in Fig.~\ref{fig:testbedtopo}. Each interferer (802.11abg, Atheros AR5213A) was sending broadcast packets of size 150\,Bytes at a rate of 200\,Hz using a PHY bitrate of 6\,Mbps (802.11a). Thus, each interferer created a channel load of 4\%. Note, that carrier sensing was activated. From Fig.~\ref{fig:channel_load}, we can learn that similar to channel 6, the channel load is unevenly distributed among the nodes. The objective is to emulate the external interference from channel 6.%\markus{Warum will man channel 6 interference simulieren? Warum sind die Scenarien ueberhaupt so gewaehlt. Bekommt man anders keine 5GHz interference? Generell sollte die wahl der Scenarien besser begruendet werden.}

%Third, we again use channel 44, but this time we self-induced interference. The location of our 2 interferer nodes is marked in Fig.~\ref{fig:testbedtopo} as A and B. Both nodes were transmitting 150 bytes MAC broadcast packets at 6\,Mbps at a rate of 600 packets per second which translates in a channel utilization of 16\% for each interferer. Note, that each interferer still conducts carrier sensing before using the medium (CSMA).

\vspace{2.5 mm}

\subsubsection*{Parameters}
The following MCS combinations were evaluated (Table~\ref{tab:mcstab}): 6 Mbps (QPSK1/2, 802.11g/a), MCSIdx=0 (QPSK1/2, 1 spatial stream), MCSIdx=7 (64QAM5/6, 1 spatial stream), MCSIdx=8 (QPSK1/2, 2 spatial streams) and MCSIdx=15 (64QAM5/6, 2 spatial streams).
Note, that for all MCS combinations with an index larger 7 spatial multiplexing is used instead of SD, i.e. no spatial diversity is utilized at the PHY layer. %Note, that STBC is only used for 802.11n with single spatial streams, i.e. no SM. 
Furthermore, the channel width was varied between 20 and 40\,MHz, an OFDM Long Guard Interval (LGI), and a packet size of 2200\,Bytes were used.

%\begin{sideways}MCS\newlineIndex\end{sideways}
\vspace*{0.5cm}
%\begin{center}
\begin{table}[h]
\begin{center}
\begin{tabular}{|>{\centering}p{.05\textwidth}|>{\centering}p{.05\textwidth}|>{\centering}p{.1\textwidth}|>{\centering}p{.1\textwidth}|>{\centering}p{.1\textwidth}|c|}
\hline
 \multirow{2}{*}{\begin{sideways}\footnotesize{MCSIdx\;\;}\end{sideways}} &
 \multirow{2}{*}{\begin{sideways}\footnotesize{\#streams\;\;}\end{sideways}} &
 \multicolumn{4}{c|}{Data rate (Mbit/s)} \\ \cline{3-6}

   & & \multicolumn{2}{c|}{20\,{MHz} channel} & \multicolumn{2}{c|}{40\,{MHz} channel} \\ \cline{3-6}
   & & LGI & SGI & LGI & \;\;\;\;\;\;\;SGI\;\;\;\;\;\;\;  \\
& & & & &        \\ \cline{1-6}
0 & 1 & \cellcolor{lightgray}{6.50} & 7.20 & \cellcolor{lightgray}{13.50} & 15.00        \\ \cline{1-6}
1 & 1 & 13.00 & 14.40 & 27.00 & 30.00      \\ \cline{1-6}
2 & 1 & 19.50 & 21.70 & 40.50 & 45.00      \\ \cline{1-6}
3 & 1 & 26.00 & 28.90 & 54.00 & 60.00      \\ \cline{1-6}
4 & 1 & 39.00 & 43.30 & 81.00 & 90.00      \\ \cline{1-6}
5 & 1 & 52.00 & 57.80 & 108.00 & 120.00    \\ \cline{1-6}
6 & 1 & 58.50 & 65.00 & 121.50 & 135.00    \\ \cline{1-6}
7 & 1 & \cellcolor{lightgray}{65.00} & 72.20 & \cellcolor{lightgray}{135.00} & 150.00    \\ \cline{1-6}
8 & 2 & \cellcolor{lightgray}{13.00} & 14.40 & \cellcolor{lightgray}{27.00} & 30.00      \\ \cline{1-6}
9 & 2 & 26.00 & 28.90 & 54.00 & 60.00      \\ \cline{1-6}
10 & 2 & 39.00 & 43.30 & 91.00 & 90.00      \\ \cline{1-6}
11 & 2 & 52.00 & 57.80 & 108.00 & 120.00      \\ \cline{1-6}
12 & 2 & 78.00 & 86.70 & 162.00 & 180.00      \\ \cline{1-6}
13 & 2 & 104.00 & 115.60 & 216.00 & 240.00      \\ \cline{1-6}
14 & 2 & 117.00 & 130.00 & 243.00 & 270.00      \\ \cline{1-6}
15 & 2 & \cellcolor{lightgray}{130.00} & 144.40 & \cellcolor{lightgray}{270.00} & 300.00 \\ \cline{1-6}
\end{tabular}
\caption{Relationship between MCS index, guard interval, bandwidth and the corresponding data rate.}
\label{tab:mcstab}
\end{center}
\end{table}
%\end{center}

%
% Results
%
\subsection{Results}

\vspace{2.5 mm}
%
% Link length
%
\subsubsection{Link length}

\begin{figure}[t]
  \begin{center}
  \mbox
    {
      \subfigure[~Scenario 1 (channel 6)]
      {
        \label{fig:link_dist24}
        \scalebox{1}
        {
            \includegraphics[width=0.35\linewidth]{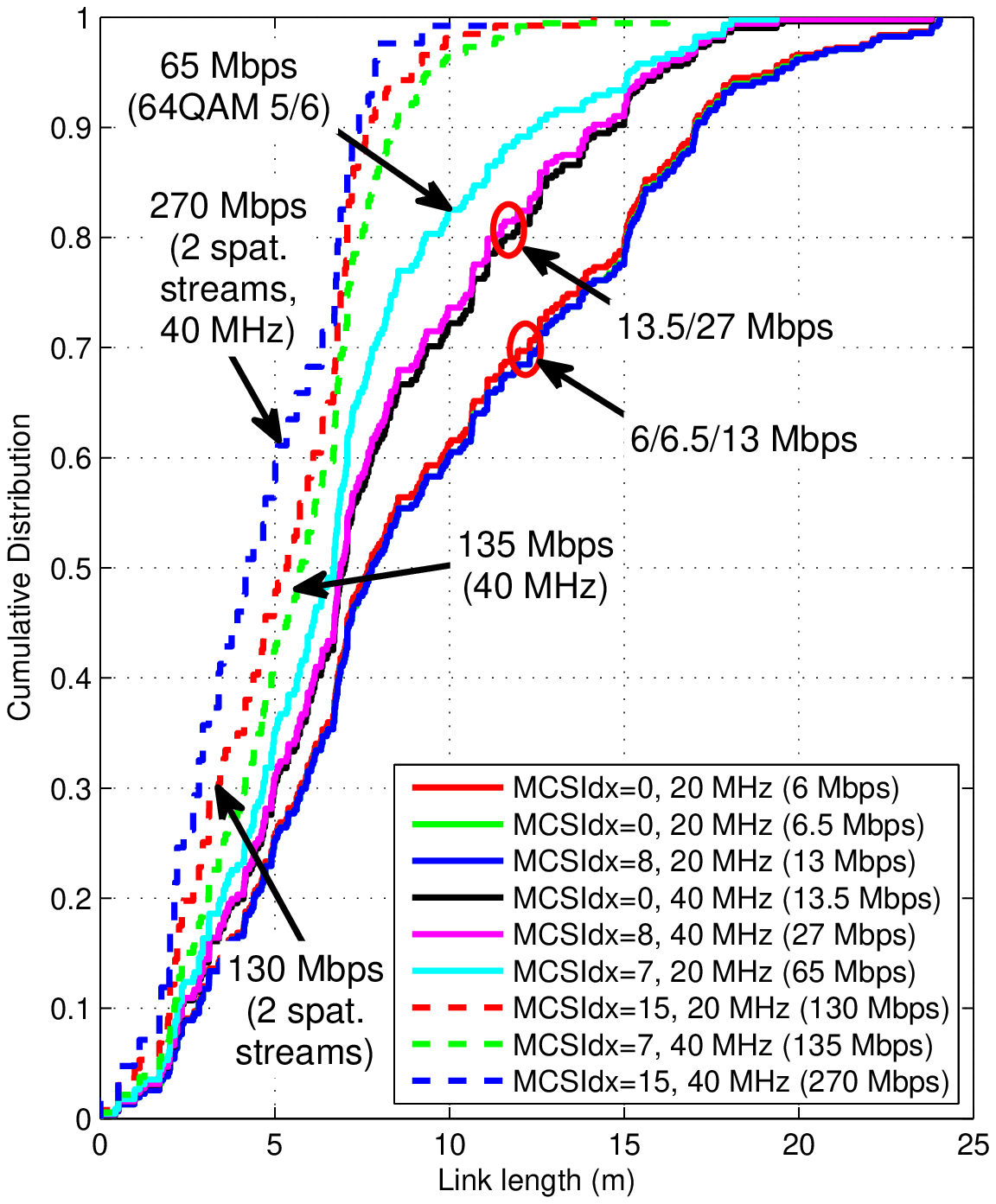}
        }
      }
%    } \\
%  \mbox
%    {      
      \subfigure[~Scenario 2/3 (channel 161)]
      {
        \label{fig:link_dist5}
        \scalebox{1}
        {
            \includegraphics[width=0.35\linewidth]{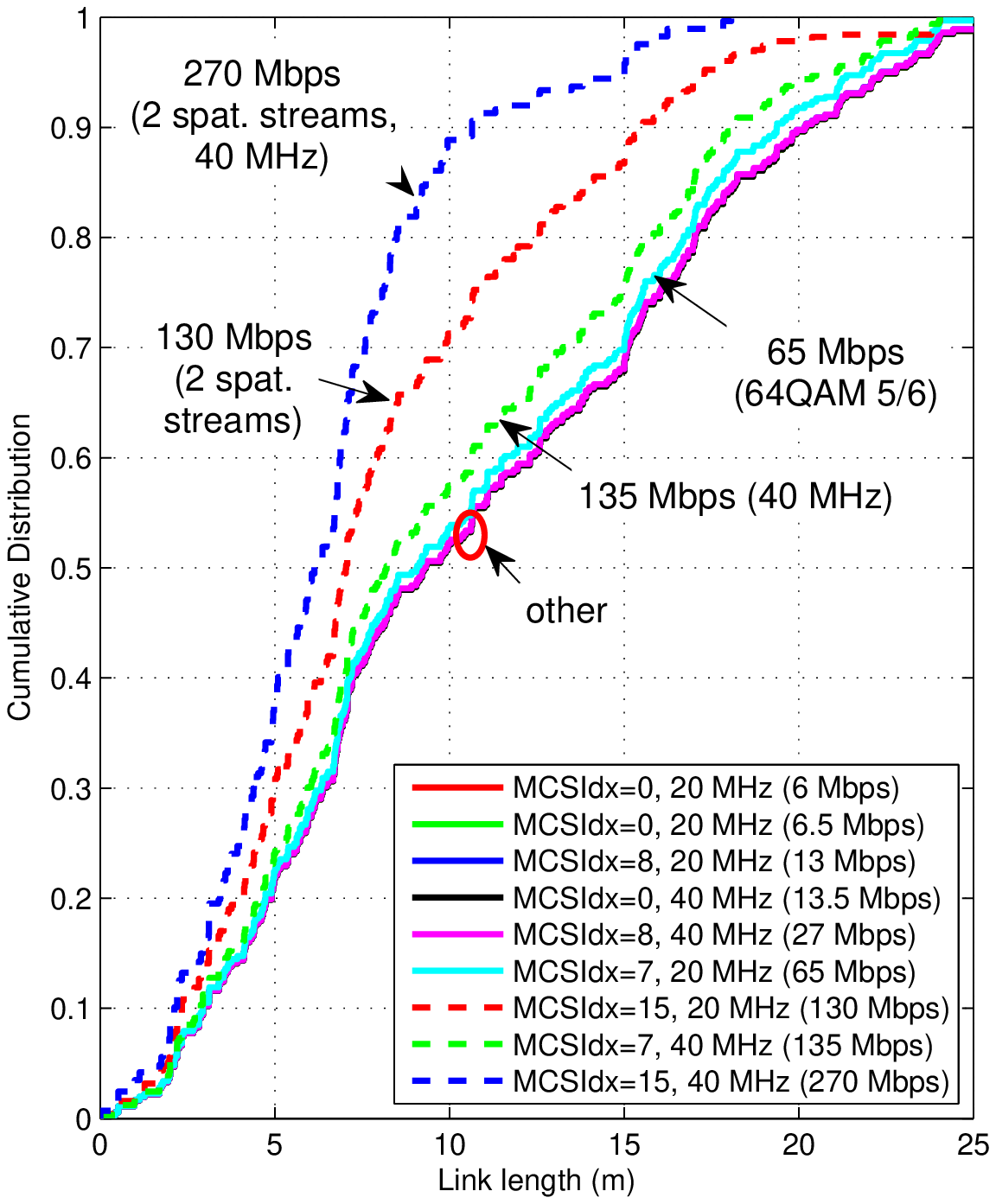}
        }
      }
    }
   \caption{Link length distribution.}
   \label{fig:link_dist}
   \end{center}
   %\markus{Generell waere es besser die links nach den verschiedenen MIMO/nicht MIMO/a,g modi aufzuschluesseln als nach der bitrate.}
\end{figure}

Before we present the actual experiment results, we need to understand the potential impact of the chosen scenario on packet losses.
We start with an analysis of link length distribution (Fig.~\ref{fig:link_dist}). In our model a link between two nodes exists when the packet error rate is below 90\%.

If compared to channel 161 (in scenario 2 and 3), links are shorter on channel 6 (scenario 1). With the lowest MCS (i.e. 6 or 6.5\,Mbps) only 10\% of the channel 6 links are longer than 17\,m, compared to 20\,m for channel 161 links. If we use the highest MCS, i.e. 270\,Mbps, the difference is even higher, i.e. the maximum link lengths are around 7.2\,m and 10.6\,m respectively. 

The short communication range, especially at high MCSs, was one reason for us to place a significant number of nodes very close to each other (ref. to house 2, fourth floor, Fig.~\ref{fig:testbedtopo}). A 5\,GHz channel on the other hand has the potential for longer links due to higher transmission power~\cite{Zubow2011}. To accustom potentially longer links on channel 161, we also placed a few nodes at longer distances. On both channels, links are shorter when using a higher MCS. 

We cannot observe any link length differences on channel 6 between 13.5 and 27\,Mbps, and between 6, 6.5, and 13\,Mbps. 
Finally, we cannot see any difference between 802.11n and the outdated 802.11g/a, if we use the lowest MCS (cmp. 6.5\,Mbps vs. 6\,Mbps). 
The improved STBC based diversity in 802.11n did not result in a notable increased communication range. Note, that the manually induced interference on channel 161 (scenario 3) had no impact on the distribution of the link length.

\vspace{2.5 mm}

%
% Link PDR
%
\subsubsection{Link Packet Delivery Ratio}

In a next step, we look at packet loss. We calculated the \emph{Packet Delivery Ratio} (PDR) for each link and each evaluated PHY mode, i.e. MCS, MIMO mode (SD vs. SM), and channel width (20 vs. 40\,MHz) combination.
Notice, MAC diversity gains are higher for environments where the majority of links have weak or intermediate PDRs. The gain from MD is low or nonexistent if the majority of links have a high PDR, i.e. $\geq 0.9$. Since MD is some kind of selection diversity it can only improve the PDR of the virtual link. From the practical point of view, links with too small PDRs, i.e. $\leq 0.1$ cannot be utilized~\cite{Biswas2005}. The required coordination between candidates of an OR transmission induce a significant management overhead which can exceed the achieved MD gain. For the following, we qualify all PDRs between 0.1 and 0.9 as intermediate.

\begin{figure}[t]
  \begin{center}
  \mbox
    {
      \subfigure[~Scenario 1 (channel 6)]
      {
        \label{fig:link_pdr24}
        \scalebox{1}
        {
            \includegraphics[width=0.45\linewidth]{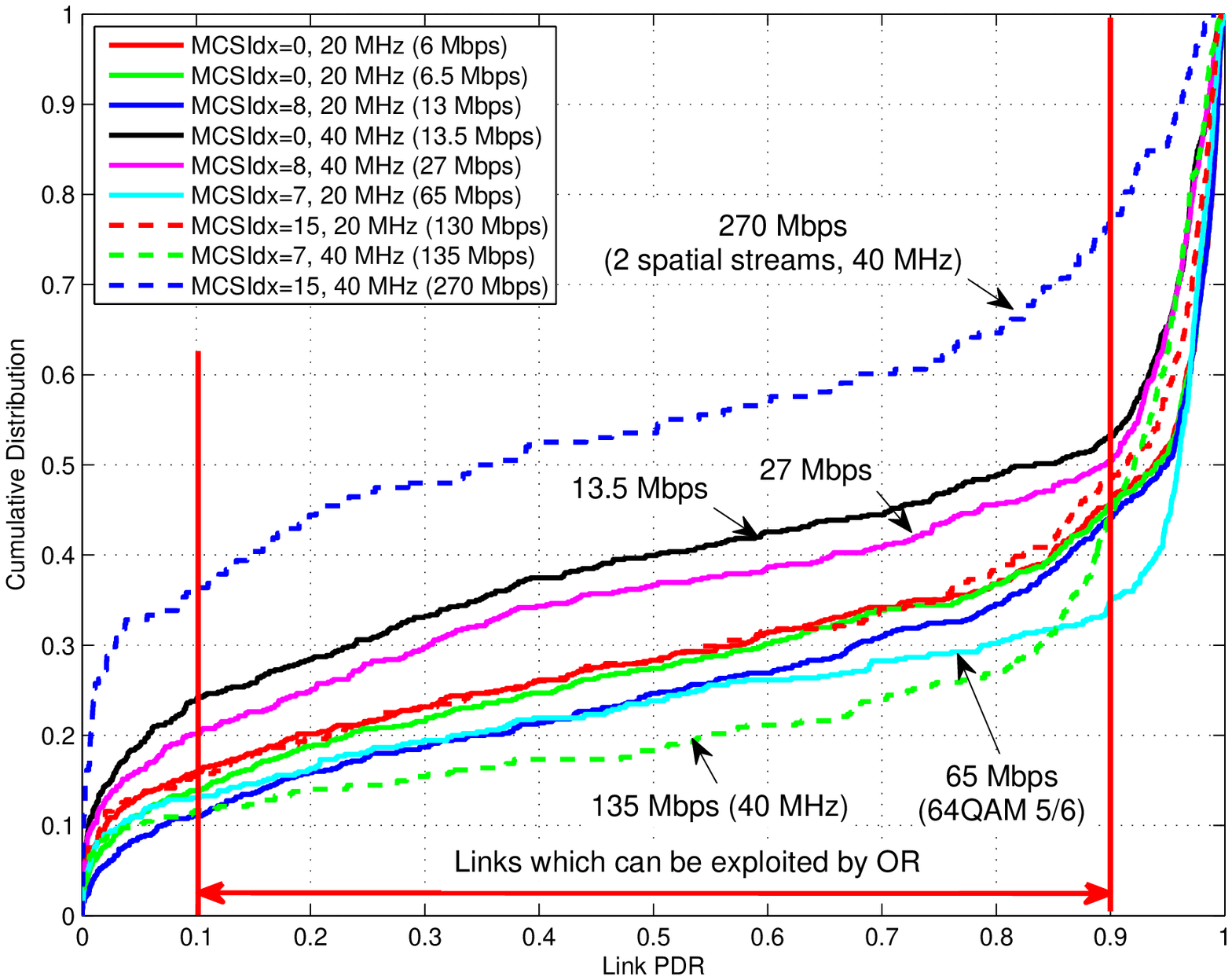}
        }
      }
%    } \\
%  \mbox
%    {        
      \subfigure[~Scenario 2 (channel 161)]
      {
        \label{fig:link_pdr5}
        \scalebox{1}
        {
            \includegraphics[width=0.45\linewidth]{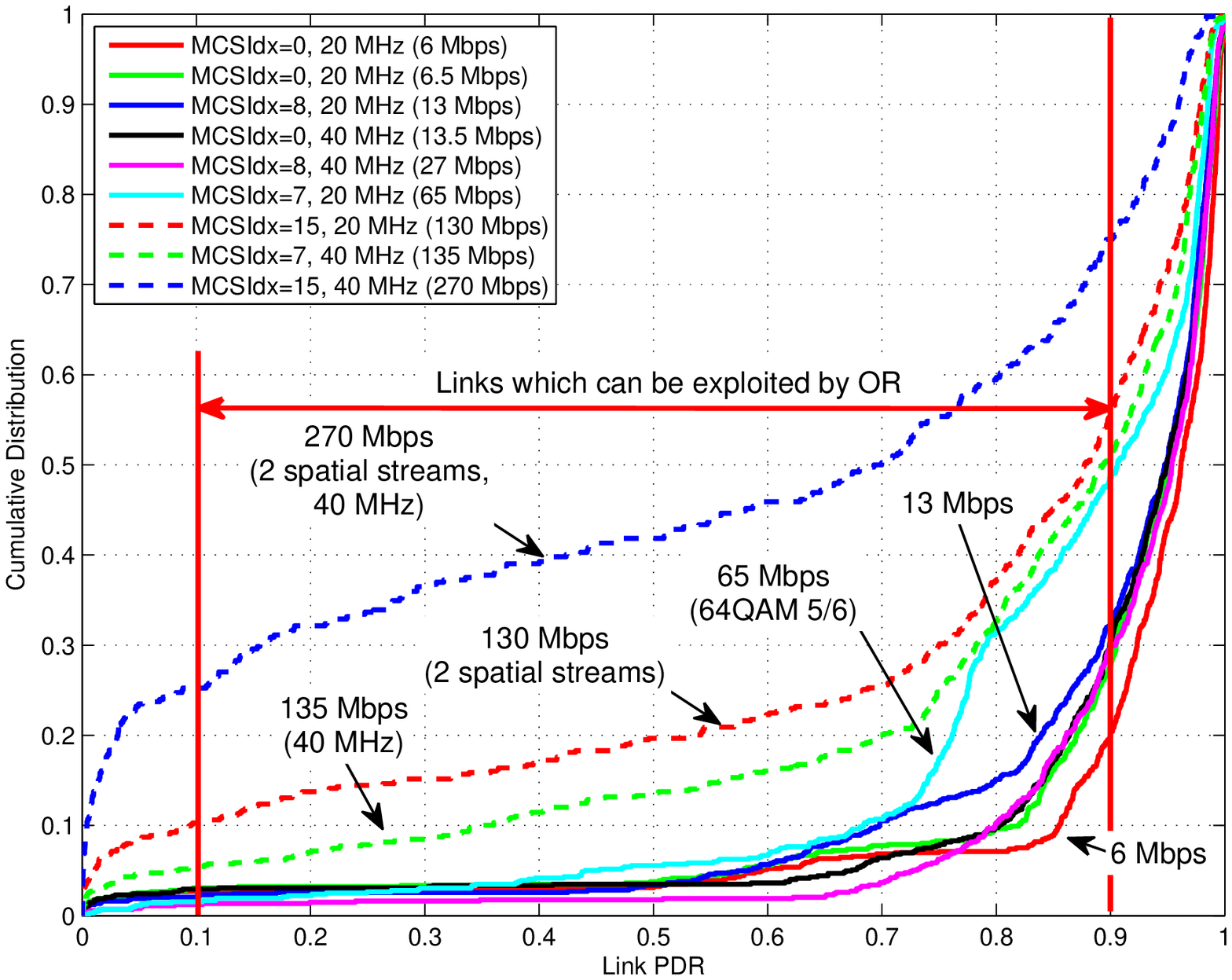}
        }
      }
    } \\
  \mbox
    {       
      \subfigure[~Scenario 3 (channel 161 + self interference)]
      {
        \label{fig:link_pdr5i}
        \scalebox{1}
        {
            \includegraphics[width=0.45\linewidth]{figs/ch161link_pdr_cdf}
        }
      }
    }
   \caption{Link Packet Delivery Ratio (PDR).}
   \label{fig:link_pdr}
   \end{center}
\end{figure}

Fig.~\ref{fig:link_pdr} shows the distribution of link PDRs for all three scenarios. In scenario 1 (channel 6), 35\% to 77\% of the links have a PDR of less than 90\% depending on the used PHY mode (Fig.~\ref{fig:link_pdr24}). We cannot identify any clear relationship between the used PHY mode and the PDR. The situation is different in scenario 2 (Fig.~\ref{fig:link_pdr5}). Here we see a clear ordering regarding the physical bitrate: with higher bitrate, we have more links with intermediate PDRs.
Nearly all links with a bitrate of 27\,Mbps or lower have a PDR of almost 1. This is different to 270\,Mbps where 55\% of the links have a PDR of less than 0.9.

Finally, in the results of experiments for the third scenario, the PDR distribution significantly changes when 15 interferer nodes are added to an otherwise empty channel. Now, links with intermediate PDRs are common, i.e. 20\% to 75\% of the links have a PDR of less than 90\% depending on the used PHY mode. Even links using a low bitrate have intermediate PDRs. 

In scenario 1, the relative number of links which can be exploited by MD (links with intermediate PDRs) is between 22\% and 40\% depending on the used PHY mode. On average only 30\% of all links can be exploited by MD, i.e. on average only every third link is suitable for MD. The situation in scenario 2 is even more inappropriate for MD. Here only 1\% to 33\% of the links have intermediate PDRs depending on the used PHY mode. The average value over all links is 6.7\%, i.e. on average every 15th link is suitable for MD. The interference in scenario 3 increases the gain from MD. Here 17\% to 50\% of the links have intermediate PDRs depending on the used PHY mode. The average value over all links is 34\%, which is comparable to scenario 1.

These initial results are very deflating. They show that the expected gain from MD in presence of PHY diversity is low when using 802.11n and is also low when the network is run in 802.11g/a mode. In the latter case, the receiver seems to still make use of the multiple antennas at the receiver side by performing Maximum Ratio Combining (MRC\footnote{MRC is a technique on the receiver side which optimally combines signals from multiple receiving antennas.}).

The MD gain to be expected is lower in absence of interference (scenario 2) and is limited to high physical bitrates only where no spatial diversity is applied, i.e. SM instead of SD. This is a crucial difference to the observations made for SISO systems, where a large majority of links with intermediate PDRs was observed~\cite{Miu2005}. Manually induced interference (scenario 3) increases the number of links with intermediate PDRs. 
From the results so far, we conclude for stationary networks with robust PHY modes ($\leq$ 27\,Mbps) that packet losses at a single receiver can be attributed mainly to interference. This is different to the explanation of Miu et al.~\cite{Miu2005} which claim that packet losses at a single receiver are due to short term channel fluctuations.

%In summary, our results show clear benefits of using MAC diversity when using an interference polluted channel. In contrast to that in an interfernce-free channel the gain from AD is limited to high physical bitrates only.

\vspace{2.5 mm}

%
% Are packet losses indep?
%
\subsubsection{Independence of packet losses}

In the previous section, we learned that there are links with intermediate PDRs in an 802.11n network. The existence of links with intermediate PDRs is only one criterion to achieve a gain from MD. The packet losses at different receivers must also be independent or at least highly uncorrelated. There is no gain from MD for two receivers with dependent packet losses.

To quantify independence of packet losses, we implemented a simple algorithm that emulates an MD algorithm by combining packet receptions from two receivers to improve the overall packet delivery ratio. This approach is similar to~\cite{Shrivastava2008}.
\mbox{PDR($A \cup B$)} represent the number of broadcast transmissions that were successfully received using this algorithm, i.e. using the MD emulation. We compare this experimentally determined PDR with the expected combined PDR $1-(1-\mathrm{PDR}_A) \cdot (1-\mathrm{PDR}_B)$.

\begin{figure}[t]
  \begin{center}
  \mbox
    {
      \subfigure[~Scenario 1 (channel 6)]
      {
        \label{fig:indep_vs_union_pdr24}
        \scalebox{1}
        {
            \includegraphics[width=0.45\linewidth]{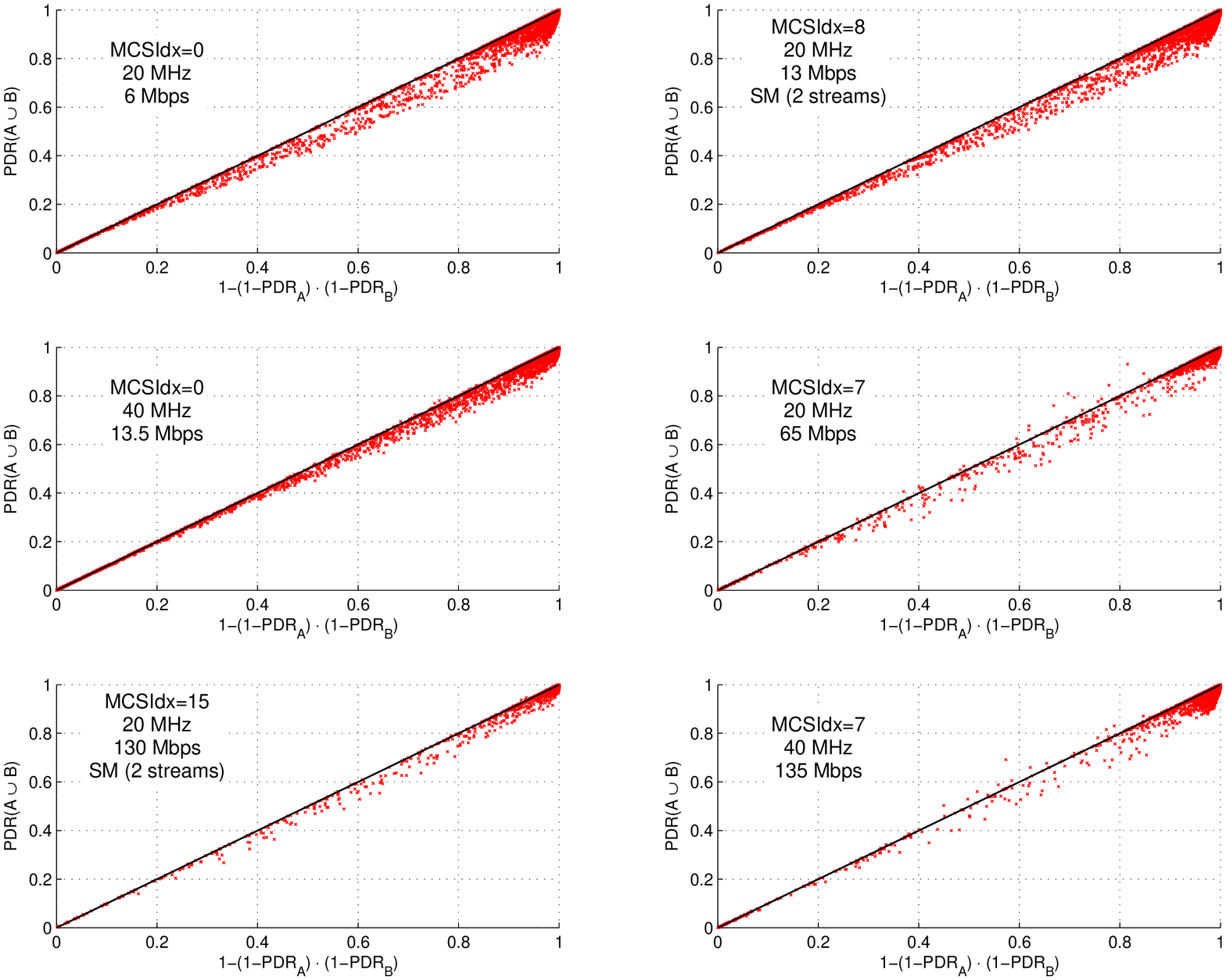}
        }
      }
%    } \\
%  \mbox
%    {       
      \subfigure[~Scenario 2 (channel 161)]
      {
        \label{fig:indep_vs_union_pdr5}
        \scalebox{1}
        {
            \includegraphics[width=0.45\linewidth]{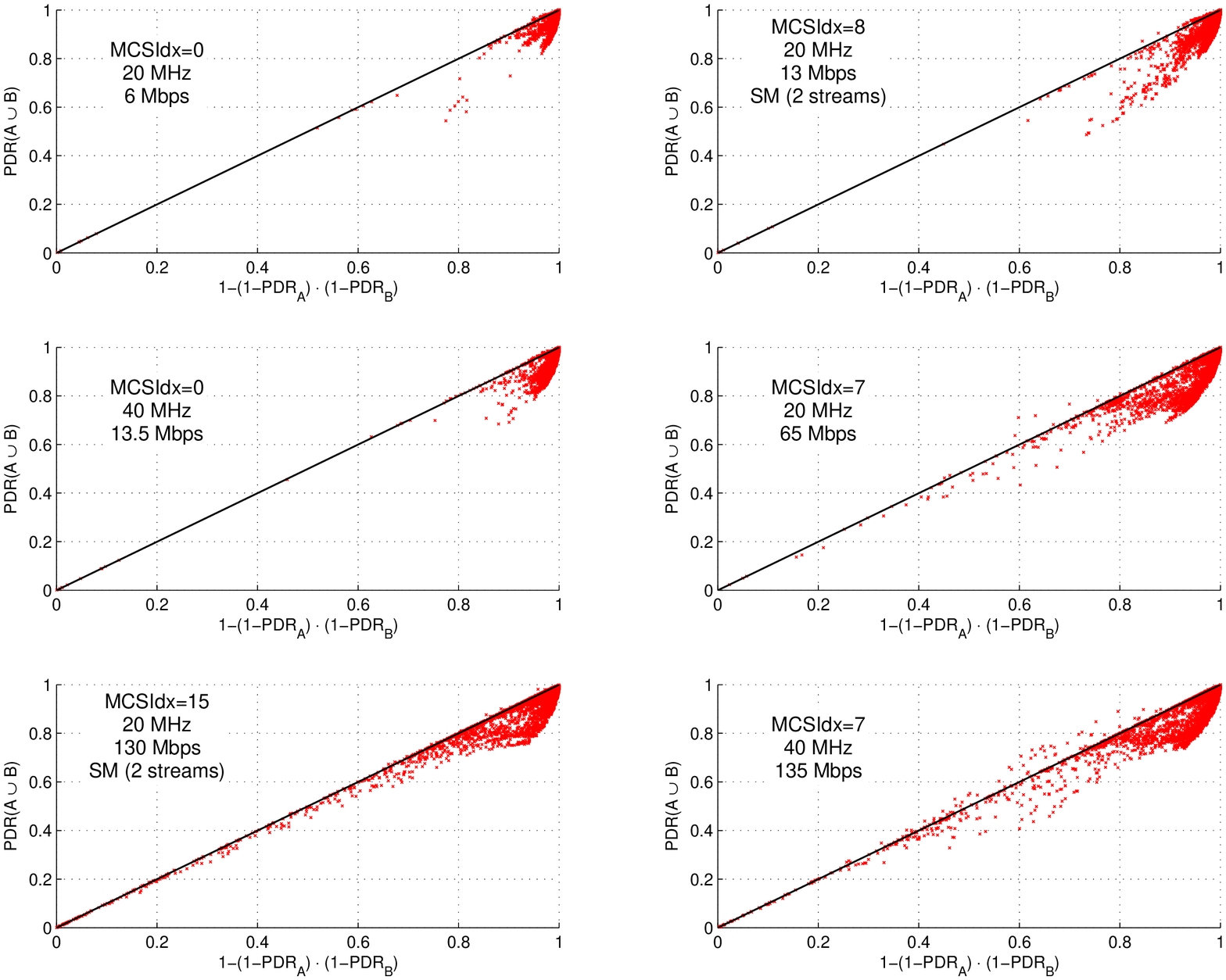}
        }
      }
    } \\
  \mbox
    {       
      \subfigure[~Scenario 3 (channel 161 + self interference)]
      {
        \label{fig:indep_vs_union_pdr5i}
        \scalebox{1}
        {
            \includegraphics[width=0.45\linewidth]{figs/ch161indep_vs_union_pdr}
        }
      }
    }
   \caption{Expected vs. actual MAC diversity gain.}
   \label{fig:indep_vs_union_pdr}
   \end{center}
\end{figure}

Fig.~\ref{fig:indep_vs_union_pdr} shows a scatter plot of both the real and expected combined PDR for all possible receiver pairs, i.e. $46 \times \binom{45}{2}$ in our case. If the packet losses at two receivers are independent, both terms are equal and thus all the points in the scatter plot should lie on the diagonal line. 
However, as shown in the figure, a large part of points does not lie on the line, especially in scenario 1 and 3. This indicates that the packet losses in 802.11n as well as 802.11g/a are dependent. This is different to the observation made in~\cite{Shrivastava2008} and similar to our observations made for SISO and 802.11b/g~\cite{Zubow2008a}. 
Especially for low MCSs, the difference between real and expected PDR can be up to 10 percentage points in scenario 1 or even higher in scenario 3. This indicates that the packet loss at different receivers may be correlated under some circumstances. 
The situation is different for scenario 2. 
First, as already mentioned, intermediate PDRs are only common for high MCSs. Secondly, we also have negatively correlated receivers where the actual gain is higher than the expected one. 
The manually induced interference increases the correlation which is always a positive correlation. This means the actual gain is lower than the expected one.

\vspace{2.5 mm}

%
% We see that packet losses are sometimes not indep? What has an impact on that? Spatial seperation between both candidates?
%
\subsubsection{Do spatially co-located receivers have correlated PDRs?}

The previous section showed that packet losses of different receivers can be dependent. In our previous work~\cite{Zubow2008a} we evaluated this for SISO systems based on 802.11b/g. 
We discovered that PDRs of physically close receivers (less than two meters distance) are correlated. This means the probability of multiple link failures can no longer be calculated by simply multiplying error rates.

\begin{figure}[t]
  \begin{center}
  \mbox
    {
      \subfigure[~Scenario 1 (channel 6)]
      {
        \label{fig:indep_vs_union_pdr_spatial24}
        \scalebox{1}
        {
            \includegraphics[width=0.45\linewidth]{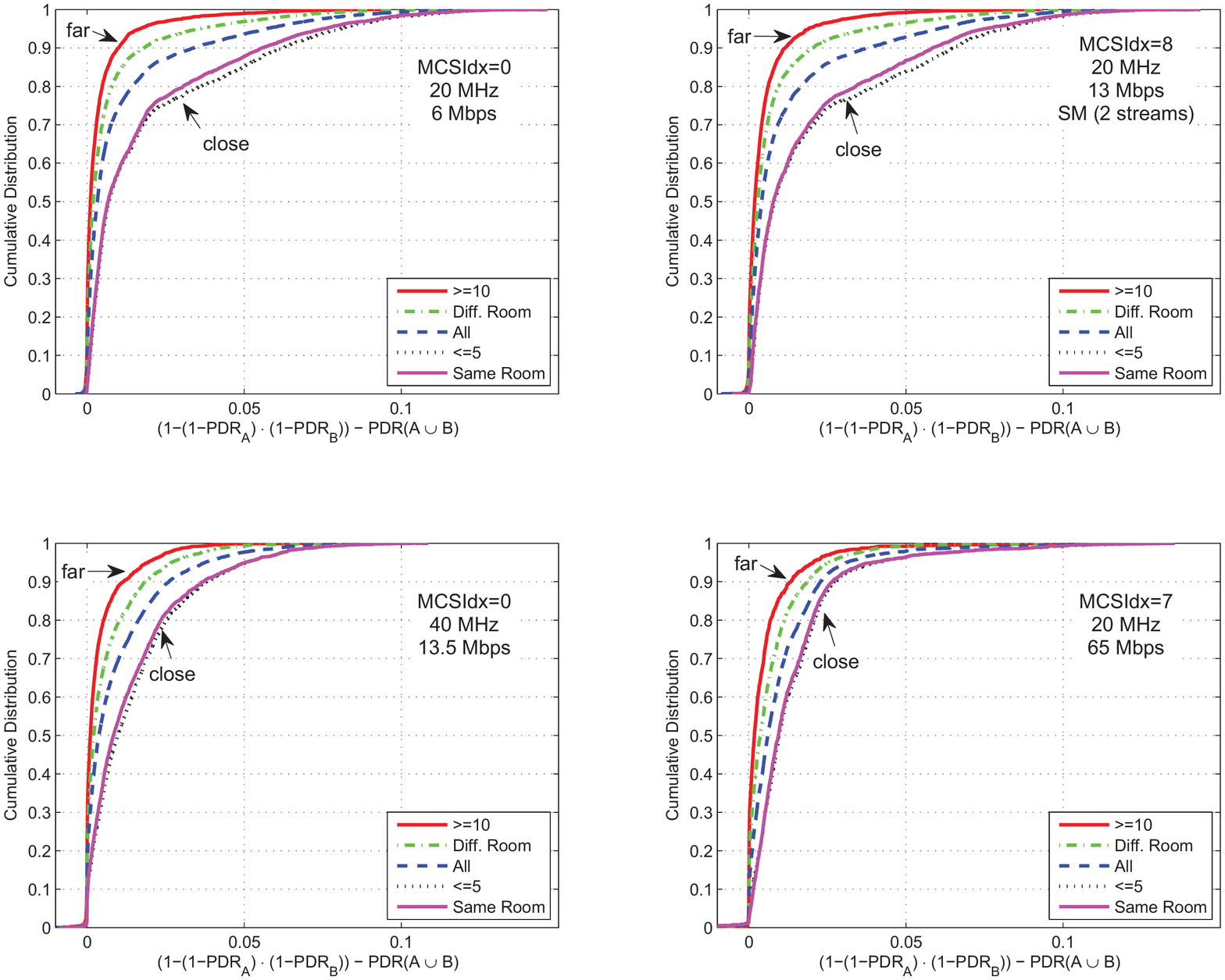}
        }
      }
%    } \\
%  \mbox
%    {      
      \subfigure[~Scenario 2 (channel 161)]
      {
        \label{fig:indep_vs_union_pdr_spatial5}
        \scalebox{1}
        {
          \includegraphics[width=0.45\linewidth]{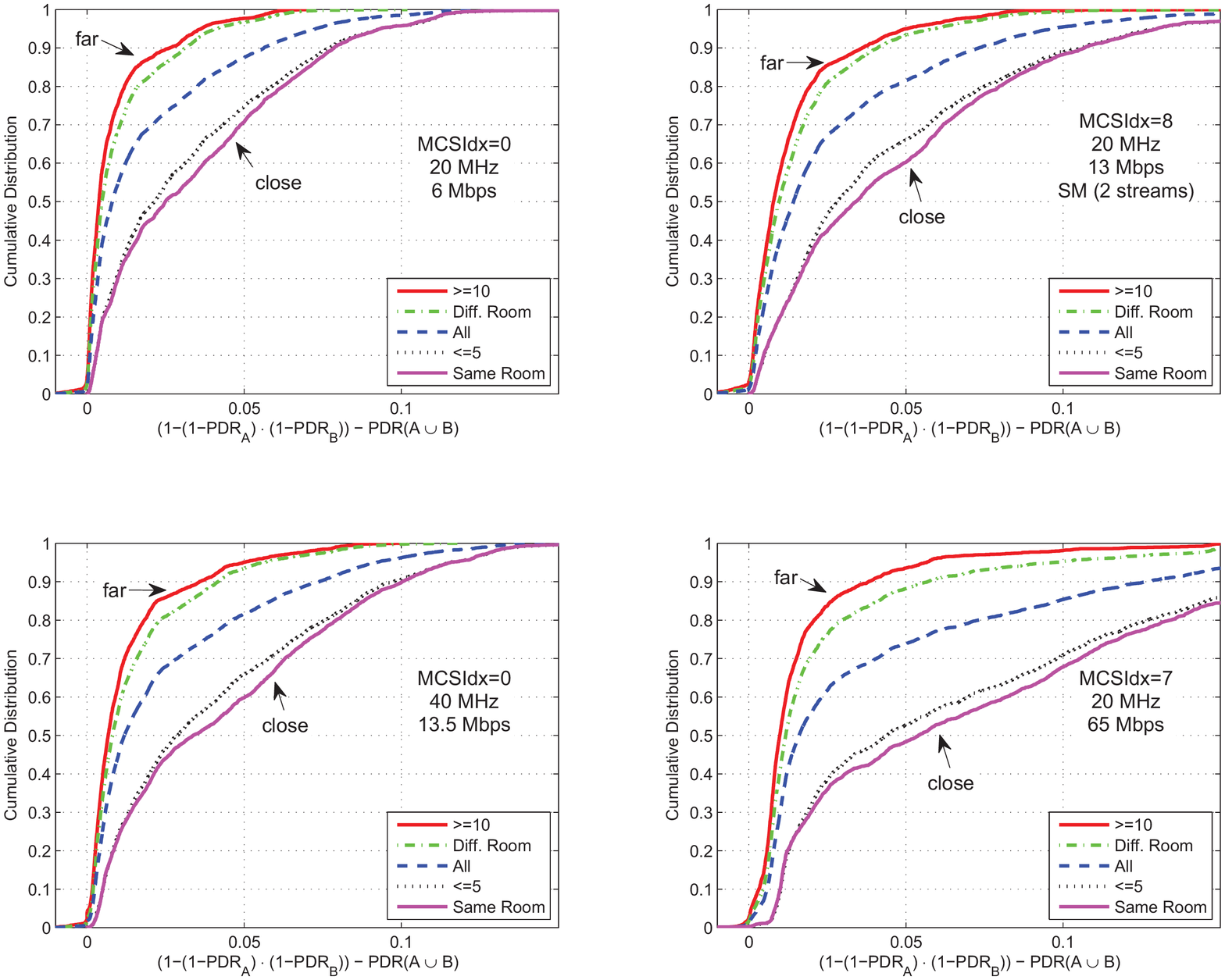}
        }
      }
    } \\
  \mbox
    {      
      \subfigure[~Scenario 3 (channel 161 + self interference)]
      {
        \label{fig:indep_vs_union_pdr_spatial5i}
        \scalebox{1}
        {
            \includegraphics[width=0.45\linewidth]{figs/ch161indep_vs_union_pdr_spatial}
        }
      }
    }
   \caption{Impact of spatial seperation between both candidates.}
   \label{fig:indep_vs_union_pdr_spatial}
   \end{center}
\end{figure}

Therefore, we compared the packet loss correlation between two receivers to the physical distance of the two receivers.
As a measure for correlation, we calculated the difference between expected PDR, $1-(1-\mathrm{PDR}_A) \cdot (1-\mathrm{PDR}_B)$ (assuming independent packet losses) and the actual PDR PDR($A \cup B$) (emulating OR). Furthermore, we classified a receiver pair according to the spatial separation between them; both receivers are either (i) in the same room or (ii) in different rooms and the Euclidean distance between both receivers is either (iii) smaller than 5\,m or (iv) larger than 10\,m. The results are shown in Fig.~\ref{fig:indep_vs_union_pdr_spatial}.

In scenario 1, the difference is larger for spatially close receivers, e.g. the difference between expected and actual MD gain of more than 1 percentage points: 10\% of the cases for high spatial separation between both receivers ($\ge 10\,m$) and 43\% of the cases for spatially close receivers (same room). The situation is even more pronounced in scenario 3 where the manually induced interference corrupts the packets of closely-located receivers. This results in highly correlated packet losses. In this case, the difference between expected and actual MD gain can be very high, e.g. up to 15 percentage points for 270\,Mbps and same room receivers. The results are different in the interference-less scenario 2. Only with 270\,Mbps, there are a very small number of correlated receiver pairs where the spatial separation between receivers has  a (only small) impact.

%\markus{Das muss besser formuliert werden und muss in die implications oder conclusions.}
%\all{
%Bottom line our results show that care must be taken when selecting the candidate set for an OR transmission especially when using high bitrates in an interference-prone channel because due to the short-range communication the candidates tend to be spatially co-located thus resulting in correlated PDRs which results in a smaller than expected gain from OR.}

\vspace{2.5 mm}

%
% What is the OR gain? Cmp. best receiver with anycast to both
%
\subsubsection{What is the gain from MD compared to choosing the best neighbor only?}

Finally, we want quantify the performance gains achievable with MD compared to choosing only the best neighbor. Therefore, we compare PDR($A \cup B$) (i.e. the rate of packets received by at least one of the receivers, i.e. as MD would receive) with the PDR of the best of the two receivers $(max(\mathrm{PDR}_A,\mathrm{PDR}_B))$. The latter represents the PDR of the best next hop as used by traditional single-path routing.

\begin{figure}[t]
  \begin{center}
  \mbox
    {
      \subfigure[~Scenario 1 (channel 6)]
      {
        \label{fig:opp_gain24}
        \scalebox{1}
        {
            \includegraphics[width=0.45\linewidth]{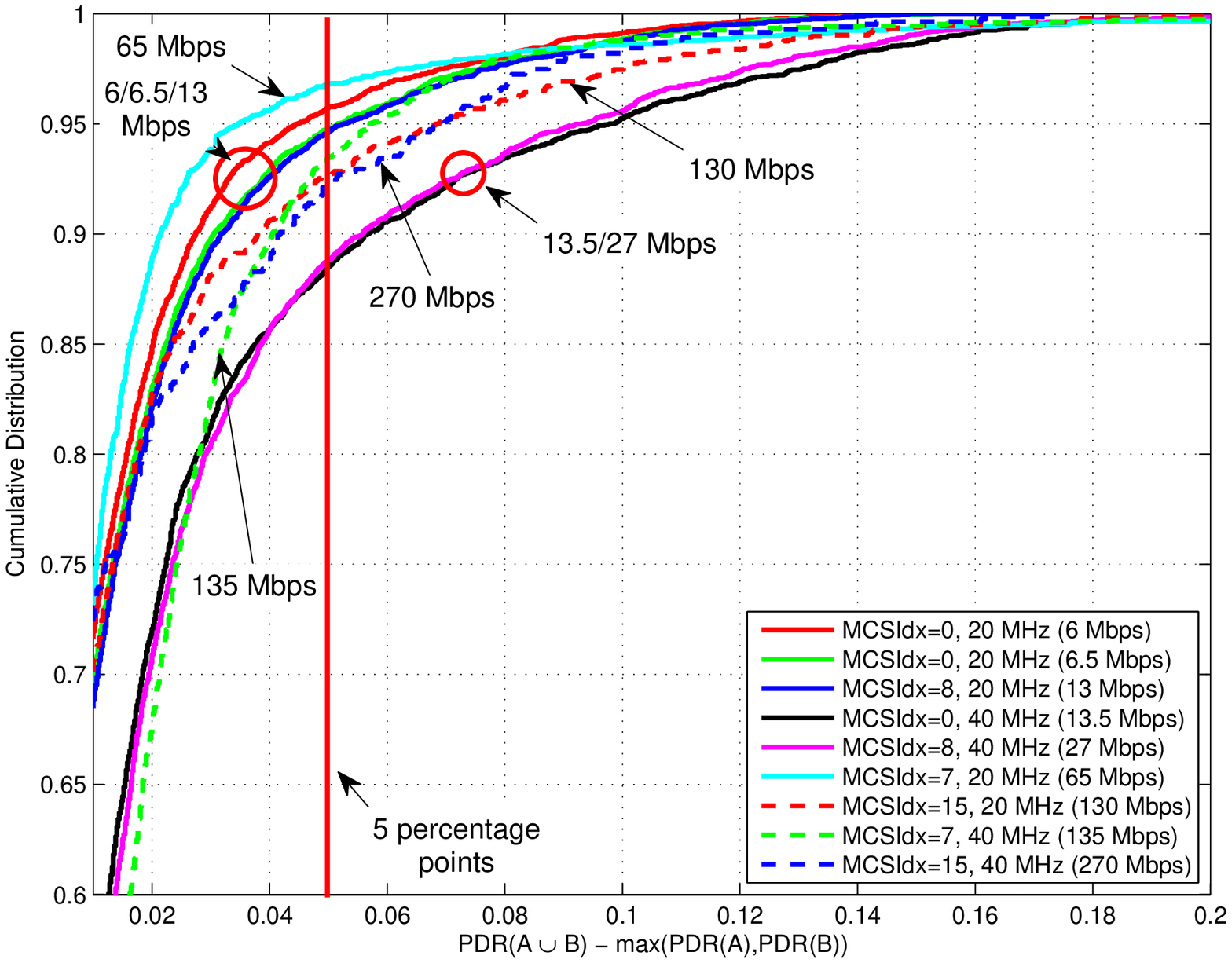}
        }
      }
%    } \\
%  \mbox
%    {       
      \subfigure[~Scenario 2 (channel 161)]
      {
        \label{fig:opp_gain5}
        \scalebox{1}
        {
            \includegraphics[width=0.45\linewidth]{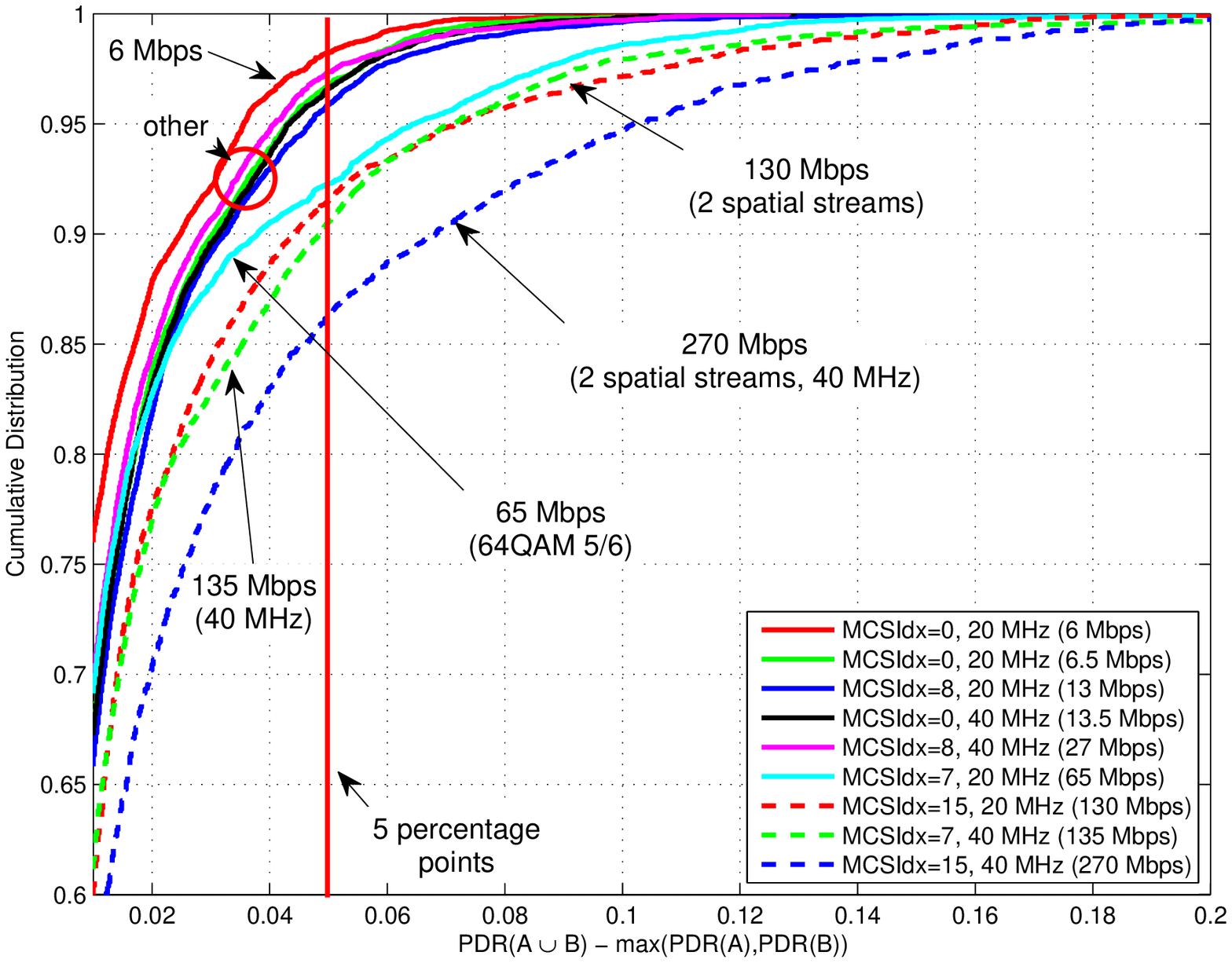}
        }
      }
    } \\
  \mbox
    {       
      \subfigure[~Scenario 3 (channel 161 + self interference)]
      {
        \label{fig:opp_gain5i}
        \scalebox{1}
        {
            \includegraphics[width=0.45\linewidth]{figs/ch161_opp_gain}
        }
      }
    }
   \caption{Gain from MD compared to choosing the best neighbor only.}
   \label{fig:opp_gain}
   \end{center}
\end{figure}

The difference between both quantities is depicted in Fig.~\ref{fig:opp_gain}. In scenario 2, the MD gain is negligible (Fig.~\ref{fig:opp_gain5}). For the highest bitrate the gain is less than 5 percentage points in 87\% of the cases, i.e. only 13\% of all evaluated receiver pairs are suitable for MD. For lower bitrates the gain is even lower, e.g. for $\leq$27\,Mbps there is no visible gain. This can be explained by the fact that scenario 2 contains only a few links with intermediate PDRs (ref. to Fig.~\ref{fig:link_pdr5}). The situation in scenario 1 is similar (Fig.~\ref{fig:opp_gain24}). Regardless of the PHY mode, less than 11\% of the receiver pairs offer an MD gain of more than 5 percentage points. The only difference is that a gain can also be achieved with low bitrates.
In scenario 3, the results are similar to scenario 1: the gain from MD is small. This means that also in the presence of interference, the gain from MD would be small.

%\markus{Das gehoehrt wieder in die implications.}
%\all{Bottom line is that when exploiting PHY diversity as provided by 802.11n regardless whether an interference-prone channel (scenario 1 and 3) or an interference-free channel (scenario 2) is used, the gain from OR as compared to TSPR is negligible. This is different to the observations made with 802.11b/g~\cite{Miu2005} and 802.11n~\cite{Shrivastava2008}.}

%\textcolor{red}{NEw: ch11: 40mhz channels=12\% of candidate sets offer a gain of 5pp or larger; 2x2=7\%, rest=$\leq$4\%. ch44: 270M=10\%; 135M=3\%; 130/65M=1.5\%; rest=0\%.}

%% file: related_work.tex
\section{Related Work}\label{sec:relatedwork}
%\markus{Maybe, the related work should be moved to the end of the paper.}\\

%In order to make a statement about the advantage from MAC diversity as provided by OR protocols the nature of packet loss need to be analyzed. 
%This section summarizes the main results from related work on the nature of packet losses.

%\tolja{Einfache link-level measurements (LLM): dabei unterscheiden zwischen 11b, 11g, 11a und 11n}

Aguayo et al.~\cite{Aguayo2004} analyzed the packet loss rates in an 802.11b outdoor mesh network. The main results were that the distribution of inter-node loss rates is relatively uniform over the whole range of loss rates. Half of the operational links in network have a loss probability higher than 30\%. 

%\tolja{Danach advanced LLM: packet loss among different receivers (indep or correlated?)}

The first study on 802.11n was from Shrivastava et al.~\cite{Shrivastava2008}. They analyzed the statistical dependence of packet losses in 802.11n receivers in a small indoor testbed. Their experimental setup was different. First, they used different hardware, i.e. Edimax (EW-7728In) 802.11n (Draft 2.0) with Ralink chipset supporting 3X3 MIMO operation and three detachable antennas. Secondly, they analyzed only the Rf polluted 2.4\,GHz band where even at night the channel utilization can be significantly high (e.g. 802.11 beacons frames) and cause interference. Thirdly, they only analyzed a single 802.11n PHY mode, i.e. Spatial Multiplexing (SM) with channel bonding SGI (300\,Mbps). Note, that in this mode MIMO is used to achieve SM and not SD which was incorrectly assumed by the authors. Furthermore, the used 40\,MHz channel is very vulnerable to interference when used in the 2.4\,GHz band. Fourthly, the two receivers representing the OR candidate set were spatially co-located with each other. The reported packet delivery ratios for both the 802.11n receivers were almost the same for almost all the locations ranging from 9\% to 80\%. Although similar loss rates were observed across both the receivers, the losses were actually independent leading to improvements in throughput due to MAC-diversity which can be exploited with OR. The reported throughput gains achieved with MD vary from 12\% to as high as 103\%. This is different to our observations.

%\tolja{Fuer Gewinn durch OR ist der nutzbare PDR-Bereich (zw. 10 und 90\%) interessant. Es reicht nicht zu sagen, nur weil 50\% der Links eine PDR kleiner 30\% haben, ist das eine Topp-Umgebung fuer OR.}
%City-wide WiFi networks, however, need to deal with poor link quality caused by urban structures and the many interferers including local WLANs. Opportunistic routing has recently emerged as a mechanism for obtaining high throughput even when links are lossy~\cite{Biswas2005}.

%% file: conclusion.tex
\section{Conclusion and Future Work}

In this paper we analyzed the gain from MAC diversity as offered by OR in the presence of physical diversity as provided by MIMO systems based on 802.11n. Therefore, we analyzed the nature of packet losses. Our experimental results obtained from an IEEE 802.11n indoor testbed show that: i) links with intermediate Packet Delivery Ratio (PDR) which can be exploited by MD are scarce, i.e. on average only 30\% and 6.7\% of all links can be exploited by MD when using an interference-prone and an interference-free channel respectively, ii) we cannot conclude that packet losses are fully independent, i.e. spatially co-located receivers have correlated PDRs which is especially the case when using an interference-prone channel. This is similar to our observations made for SISO systems based on 802.11b/g~\cite{Zubow2008a}, iii) the gain from MD is negligible regardless whether the interference-prone or an interference-free channel is used, i.e. less than 5 percentage points for PDR in 90\% of the cases compared to choosing the best neighbor only as used by traditional single-path routing. This is different to the observations made with SISO systems, e.g. 802.11b/g~\cite{Miu2005}, as well as first studies of 802.11n~\cite{Shrivastava2008}.

As future work we consider the following steps. First, we want to repeat our experiment in an outdoor environment as well as using 802.11n hardware from other vendors. Second, we want to analyze the nature of bit errors in packets with incorrect Cyclic Redundancy Check (CRC) checksum. In the past we already showed that in case of 802.11b/g the bit errors over different receiver nodes were suitably distributed, so that a correction was possible by combining OR with Network Coding techniques~\cite{Kurth2009}. Finally, we want to re-evaluate existing OR protocols using 802.11n allowing us to determine the gain from OR that is not attributed to MD.